\documentclass[prl,twocolumn,showpacs,preprintnumbers,nofootinbib]{revtex4}

\usepackage{graphicx}% Include figure files
\usepackage{dcolumn}% Align table columns on decimal point
\usepackage{bm}% bold math
\usepackage{xspace}
\usepackage{multirow}
\usepackage{amsmath}
\usepackage{amssymb}
\usepackage{hyperref}
\usepackage{aas_macros}
\usepackage{soul}
\usepackage{cancel}
\usepackage{xcolor}

\font\FermiSmallfont=cmssq8 scaled 1200

\def\LANLppthead#1{
\null 
\begin{center}\vskip -1.0truein{\hbox to 7.0truein {
\hfill
\vbox to 1in {\vfill \FermiSmallfont
              \hbox{#1}
              \vfill}
}}\vskip-0.0truein\end{center}}

\newcommand{\neff}{\ensuremath{N_{\mbox{\scriptsize eff}}}\xspace}
\newcommand{\neffth}{\ensuremath{N_{\mbox{\scriptsize eff}}^{\mbox{\scriptsize (th)}}}\xspace}
\newcommand{\neffobs}{\ensuremath{\tilde N_{\mbox{\scriptsize eff}}}\xspace}
\newcommand{\tneff}{\ensuremath{\tilde N_{\mbox{\scriptsize eff}}}\xspace}
\newcommand{\summnu}{\ensuremath{\sum m_{\mbox{\scriptsize $\nu$}}}\xspace}
\newcommand{\barnum}{\ensuremath{\omega_{\mbox{\scriptsize $b$}}}\xspace}
\newcommand{\numr}{$\nu$MR\xspace}

\begin{document}

%\preprint{APS/123-QED}
\preprint{LA-UR-14-29625}

%\title{Effects of neutrino rest mass on \neff and ionization 
%       equilibrium freeze-out}
\title{Effect of neutrino rest mass on ionization equilibrium
       freeze-out}

\author{E. Grohs$^1$}
\author{G. M. Fuller$^1$}
\author{C. T. Kishimoto$^{1,2}$}
\author{M. W. Paris$^3$}

\affiliation{$^{1}$Department of Physics, University of California,
San Diego, La Jolla, California 92093, USA}
\affiliation{$^{2}$Department of Physics, University of
San Diego, San Diego, California 92110, USA}
\affiliation{$^{3}$Theoretical Division, Los Alamos National
Laboratory, Los Alamos, New Mexico 87545, USA}

\date{\today}

\begin{abstract}
We show how small neutrino rest masses can increase the expansion rate
near the photon decoupling epoch in the early universe, causing an
earlier, higher temperature freeze-out for ionization equilibrium
compared to the massless neutrino case.  This yields a larger
free--electron fraction, thereby affecting the photon diffusion length
differently than the sound horizon at photon decoupling.  This
neutrino-mass/recombination effect depends strongly on the neutrino
rest masses. Though below current sensitivity, this effect could be
probed by next-generation cosmic microwave background experiments,
giving another observational handle on neutrino rest mass.  
\end{abstract}

\pacs{98.80.-k,95.85.Ry,14.60.Lm,26.35.+c,98.70.Vc}

\maketitle

The history of the early universe is a history of freeze-outs, where
reaction rates fall below the Hubble expansion rate. We point out here
that the energy density associated with neutrino rest mass results in
a subtle increase in the expansion rate at photon decoupling. This
causes an earlier, higher temperature epoch for the freeze-out of
ionization equilibrium. The physics of this freeze-out and its
relation to observations of the cosmic microwave background (CMB) is a
well studied issue
\cite{1999NuPhB.543..269D,1982NuPhB.209..372C,1999PhRvD..59j3502L,
   Mangano:3.040,2010JCAP...05..037S,2013PhRvD..87h3008H,
Birrell.PhysRevD.89.023008,2014ApJ...782...74H}.
The effect we
consider, easily derivable with existing CMB analysis
tools\cite{Howlett:2012mh}, has been mentioned\cite{Lesgourges:2012nm} but not
computed quantitatively. We find that these neutrino rest-mass induced changes
in CMB observables are below the sensitivity of current methods used to observe
and analyze the CMB data. The effects, however, may be within the reach of the
next generation of precision CMB observations coupled with a self-consistent
computational approach.

Here we focus on the influence of the recombination history on CMB
observables, in particular the sound horizon $r_s$ and the photon
diffusion length $r_d$.  The earlier ionization
freeze-out caused by neutrino rest mass affects the deduced radiation
energy density in a perhaps unexpected way. 
The quantities $r_s$ and $r_d$ are given in terms
of integrals over the scale factor $a$ \cite{2013arXiv1303.5076P}:
\begin{align} 
\label{eqn:rs}
r_s   &= \int_0^{a_{\gamma d}}\frac{da}{a^2H}
         \frac{1}{\sqrt{3(1+R)}}, \\
\label{eqn:rd}
r_d^2 &= \pi^2\int_0^{a_{\gamma d}}\frac{da}{a^2H}
         \frac{1}{an_e(a)\sigma_T}
         \frac{R^2+\frac{16}{15}(1+R)}{6(1+R)^2},
\end{align} 
where $H=H(a)$ is the Hubble expansion rate, $\sigma_T$ is the Thomson
cross section, $n_e(a)$ is the free--electron number density, and
$R(a)\equiv3\rho_b/(4\rho_\gamma)$ is a ratio involving the baryon
rest mass and photon energy densities, $\rho_b$ and $\rho_\gamma$,
respectively. The integrals span the early history of the universe,
ending at $a_{\gamma d}$, the epoch of photon decoupling at a redshift
$z = 1090.43$ \cite{2013arXiv1303.5076P}.  In the analysis to follow,
we ignore the small dependence of the value of $a_{\gamma d}$ on
\summnu.\footnote{This is not entirely self-consistent but we will 
   demonstrate
   that a consistent treatment changes the decoupling redshift from
   $z=1090$ to $z=1091$.  The associated change in $a_{\gamma d}$ has
   negligible effect on $r_s$ and $r_d$; this is similar to the
finding in Ref.\ \cite{2013PhRvD..87h3008H}.}
We should note that Eq.\ \eqref{eqn:rd} is approximate and a more
complete analysis would include effects beyond the tight-coupling
approximation\cite{Hu.1996ApJ.471.542H}. 

\begin{figure}[b]
\includegraphics[width=\columnwidth]{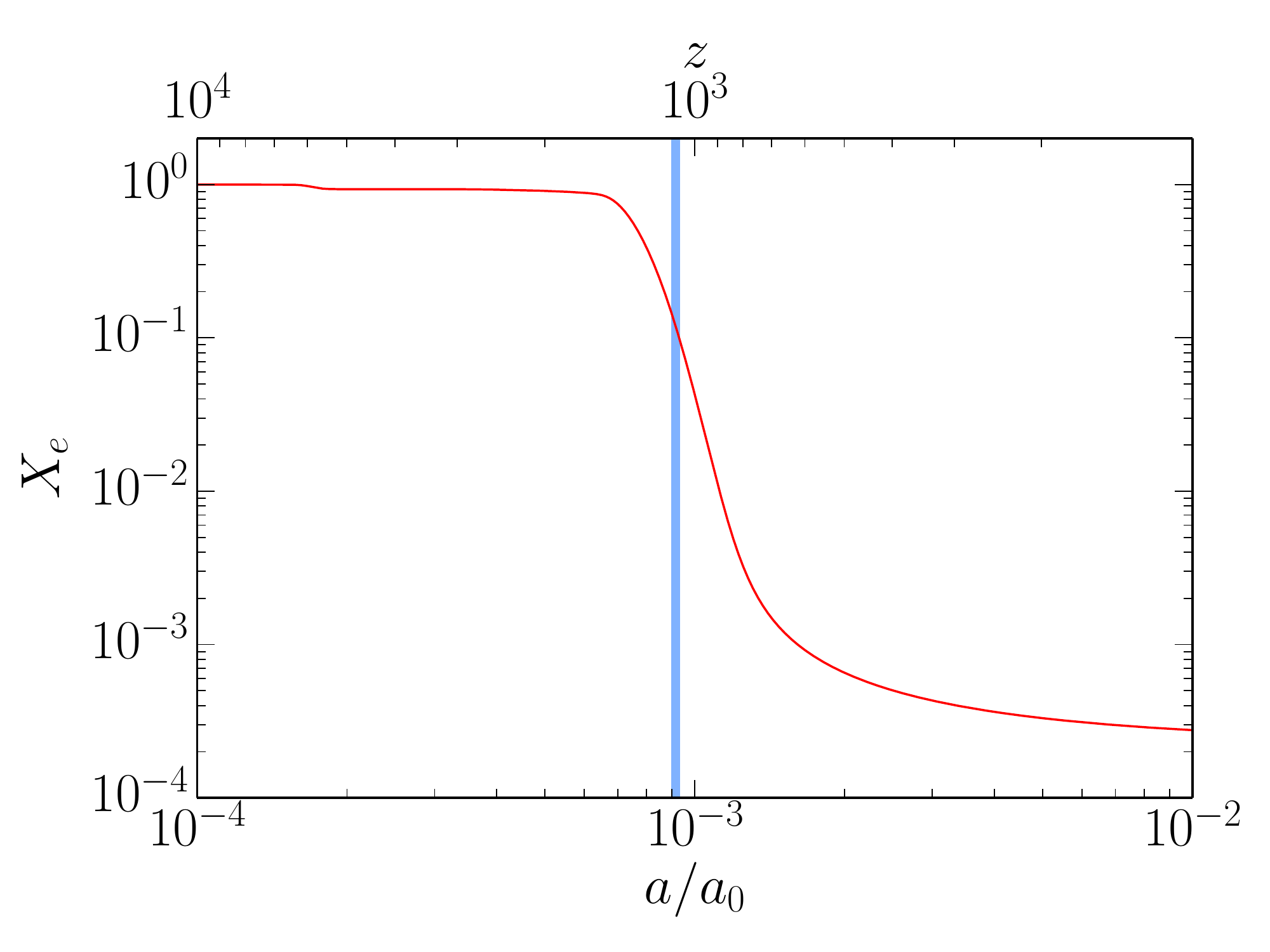}
\caption{\label{fig:xe_vs_a_0}(Color online.) The free--electron
   fraction, $X_e$, is given as a function of scale factor ratio,
   $a/a_0$ ($\equiv1$ at current epoch), and redshift, $z$, (at top).
   The primordial helium mass fraction is taken to be
   $Y_P = 0.242$. Photon decoupling, denoted by the
   shaded, vertical bar, corresponds to the epoch
   as determined by Ref.\ \cite{2013arXiv1303.5076P}.}
\end{figure}

The photon diffusion length $r_d$ depends on the number density of free
electrons $n_e^{\rm (free)}$.  The free--electron fraction, $X_e\equiv n_e^{\rm
(free)}/n_e^{\rm (total)}$, parameterizes the free--electron number density. 
%Although electrons are indistinguishable and have no predilection to recombine
%onto either hydrogen or helium, we use Eq.\ \eqref{eq:free_e_def} to
%differentiate the contributions of each ion in our recombination network
%{\color{red}, namely H II, He II, and He III.}
To evolve $X_e$, our recombination network uses the Saha equation to treat
recombination onto He III and coupled Boltzmann equations
\cite{1968ApJ...153....1P,1968ZhETF..55..278Z} to treat other recombination and
ionization processes associated with H II, He II, and He III.  We employ a
recombination reaction network which is similar to, but independent of, the
code \texttt{recfast}\cite{1999ApJ...523L...1S}.  Figure \ref{fig:xe_vs_a_0}
shows a calculation of the free--electron fraction as a function of scale factor
ratio $a/a_0$ ($\equiv1$ at current epoch), where we have taken $\summnu=0$.
Effects due to reionization processes at low redshift, $z \sim
\mathcal{O}(1)$ are neglected.  
The free--electron fraction of Fig.\ref{fig:xe_vs_a_0}, evolved through the
photon decoupling epoch, shows the freeze-out from ionization equilibrium. The
first drop from the initial value of $X_e=1$ near $a/a_0 \simeq 2\times
10^{-4}$ is a consequence of the recombination onto He III.

\begin{figure}[t]
\includegraphics[width=0.5\textwidth]{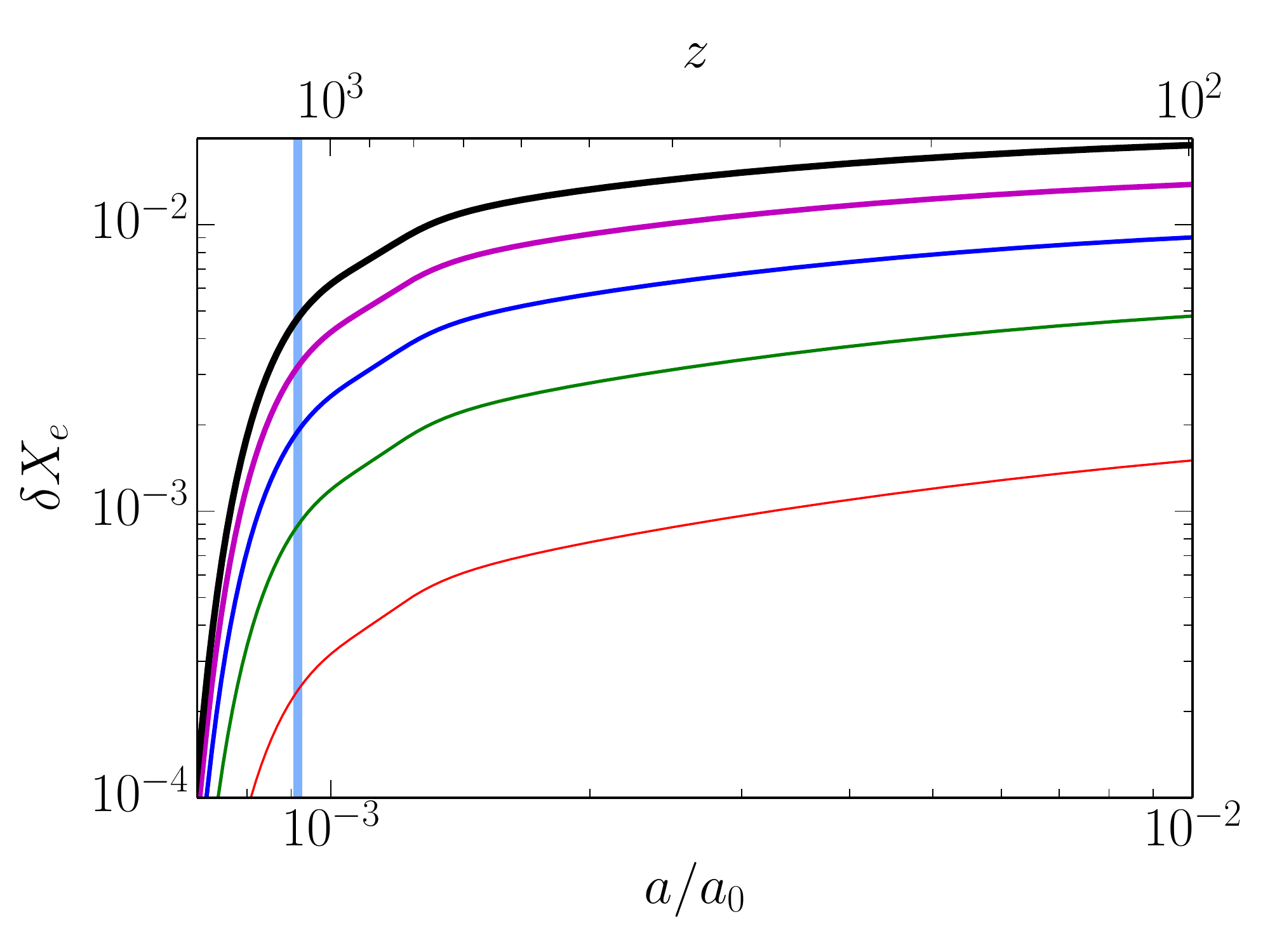}
\caption{\label{fig:xe_vs_a_non}(Color online.) The relative change in
   the free--electron fraction, $\delta X_e=\Delta X_e/X_e$ given as a function
   of scale factor ratio, $a/a_0$, and redshift, $z$, (at top).  The
   primordial helium mass fraction and vertical bar are identical to
   Fig.\ \ref{fig:xe_vs_a_0}.  Each curve corresponds to a different
   non-zero \summnu.  The curves are in equal increments of
   $\Delta\Sigma m_\nu = 0.2$ eV, starting with the smallest change
   for $\summnu = 0.2$ eV and ending with the largest change for
$\summnu = 1.0$ eV.}
\end{figure}

In Fig.\ \ref{fig:xe_vs_a_non} we plot the change in $X_e$ for non-zero values
of \summnu relative to the case with $\summnu=0$.  Non-zero \summnu has a
discernible effect on the freeze-out of $X_e$.  A larger \summnu implies a
larger Hubble rate giving an earlier epoch for $X_e$ freeze-out.
%In a study of the expansion rate during recombination,
%Ref.\cite{Zahn.PhysRevD.67.063002} observed that
%scaling the Hubble rate affects the recombination history; they didn't
%explicitly consider the role of massive neutrinos in recombination.
In a study of the expansion rate during recombination,
Ref.\ \cite{Zahn.PhysRevD.67.063002} observed that scaling the Hubble
rate affects the recombination history.  Here we build on this
argument to explicitly consider the role of neutrino rest mass on
recombination.  The curve describing the largest change corresponds to
$\summnu = 1.0$ eV, whereas the smallest change corresponds to
$\summnu = 0.2$ eV; consecutive curves are spaced by $\Delta\Sigma
m_\nu = 0.2$ eV.  

Considering Eqs.\ \eqref{eqn:rs} and \eqref{eqn:rd} we note that both
quantities, $r_s$ and $r_d$ depend on the expansion history $H(a)$.
Only the diffusion length, however, depends explicitly on the
recombination history $n_e(a)$, which is itself dependent on the expansion
history. This recombination effect changes $r_d$ and this change is 
opposite to the effect of the change of that due directly to the 
Hubble expansion.

A scaling analysis, similar to that of
Ref.\ \cite{2013PhRvD..87h3008H}, demonstrates the approximate
relation between the sound horizon, the diffusion length, and the
Hubble rate.  Consider a scale transformation to the Hubble rate, $H
\to \lambda H$, and the corresponding alteration to $r_s$ and $r_d$.
If we neglect the dependence of $R(a)$ and $n_e(a)$ on $\lambda$ we
have
\begin{align}
\label{eqn:scaling} 
r_s \propto\frac{1}{\lambda}  \text{ and }
r_d \propto\frac{1}{\sqrt{\lambda}} &
\implies\frac{r_s}{r_d}\propto\frac{1}{\sqrt{\lambda}}.
\end{align} 
These relations suggest that a larger Hubble rate ($\lambda > 1$) results in a
smaller value of the ratio $r_s/r_d$.  However, we show
below that when the dependence of the recombination history $n_e(a)$ on the
expansion $H(a)$ is taken into account $r_s/r_d$ increases.

The radiation energy density is not directly measured by observation
of the CMB.  References \cite{2013PhRvD..87h3008H} and
\cite{2014ApJ...782...74H}, however, have shown that the ratio of the
sound horizon to the photon diffusion length at the photon decoupling
epoch is sensitive to the radiation energy density. Consequently, we
distinguish between \neffth, the input parameter that determines the
radiation energy density in Eq.\ \eqref{eqn:baby_formula}, and a measure
of radiation energy density inferred from observations of the CMB,
which we shall term \neffobs.  The theoretical definition of \neff
arises from the familiar parameterization of radiation energy density
$\rho_\text{rad}$ in terms of the photon temperature $T(a)$ at
decoupling $T_\gamma\equiv T(a_{\gamma d})$ given by:
\begin{align}
   \label{eqn:baby_formula}
   \rho_\text{rad} = \left(1 +
   \frac{7}{8}\left(\frac{4}{11}\right)^{4/3}\neffth\right)
   \frac{\pi^2}{15}T_\gamma^4.
\end{align}
We adorn \neff with a superscript $\mathrm{(th)}$ to distinguish the
theoretical version of \neff, an {\em input} parameter in public
Boltzmann codes\cite{Howlett:2012mh}, from the CMB {\em inferred}
value of \neff, \neffobs described below. Calculations which include
non-equilibrium processes in the early universe suggest
$\neffth=3.046$ \cite{1999NuPhB.543..269D,1999PhRvD..59j3502L,
Mangano:3.040,neff:3.046}.  We determine \neffobs by computing the
sound horizon, $r_s$ and the photon diffusion length, $r_d$ at the
photon decoupling epoch, as follows.

\begin{figure}
   \includegraphics[width=0.5\textwidth]{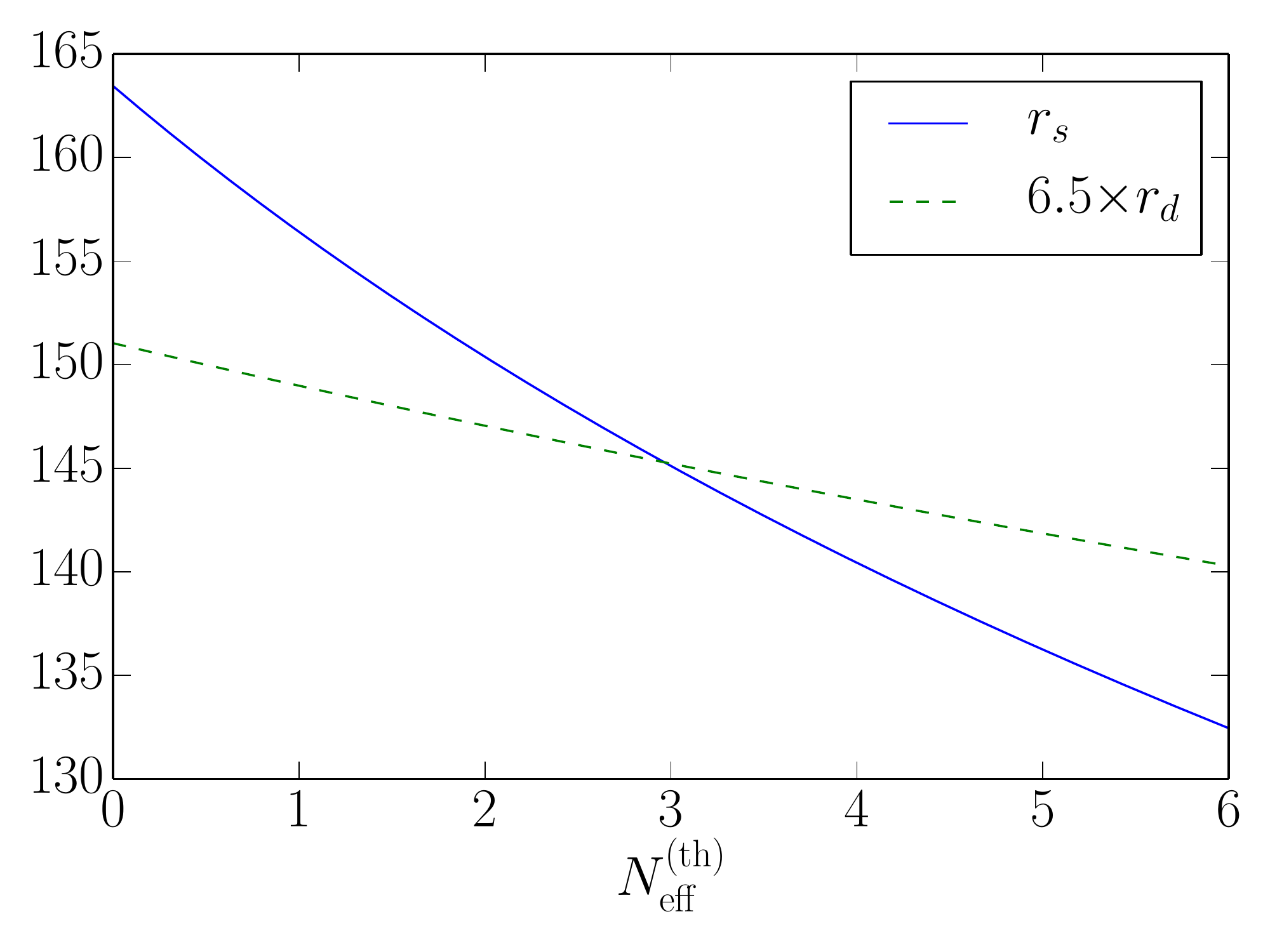}
   \caption{\label{fig:rsrdvneffth}(Color online.)
   The sound horizon $r_s$ and diffusion
   length $r_d$ as a function of \neffth.}
\end{figure}

Active neutrinos decouple from the plasma with ultra-relativistic kinematics.
As their occupation probabilities are comoving invariants thereafter, the
energy density of non-degenerate neutrinos with rest masses
$m_{\nu_i}$ and neutrino temperature $T_\nu$ is:
\begin{align}
   \rho_\nu(m_{\nu_i},T_\nu)
   &= \sum_i \int \frac{d^3p}{(2\pi)^3}\,E_i
   f_\nu(p,T_\nu)\\ 
   \label{eqn:rho_nu} 
   &= \frac{1}{2\pi^2} \sum_i \int_0^\infty dp\,p^2
   \frac{\sqrt{p^2+m_{\nu_i}^2}}{e^{p/T_\nu} + 1},
\end{align}
where the sum is over active neutrino mass eigenstates, $\nu_i$ and
the second expression follows from an assumption of decoupled
neutrinos, ignoring the small distortion due to irreversible neutrino
transport effects.  With this assumption, the neutrino energy density
for a given mass eigenstate is given by the product of the
ultra-relativistic Fermi-Dirac occupation probabilities, $f_\nu =
(\exp{(p/T_\nu)} + 1)^{-1}$, with the energy dispersion relation of a
massive particle, $E_i=\sqrt{p^2+m_{\nu_i}^2}$.
%As the rest mass is an invariant quantity, the ratio
%$m_{\nu_i}/T_{\nu}$ is a continuously changing quantity with time and
%is inconsistent with Eq.\eqref{eqn:baby_formula}. 
Therefore, the energy density in the presence of a massive-neutrino
species is larger than in the massless case and becomes increasingly
significant at later times; it does not scale as $T^4 \sim a^{-4}$ as
in Eq.\ \eqref{eqn:baby_formula}.

\begin{figure}
   \includegraphics[width=0.5\textwidth]{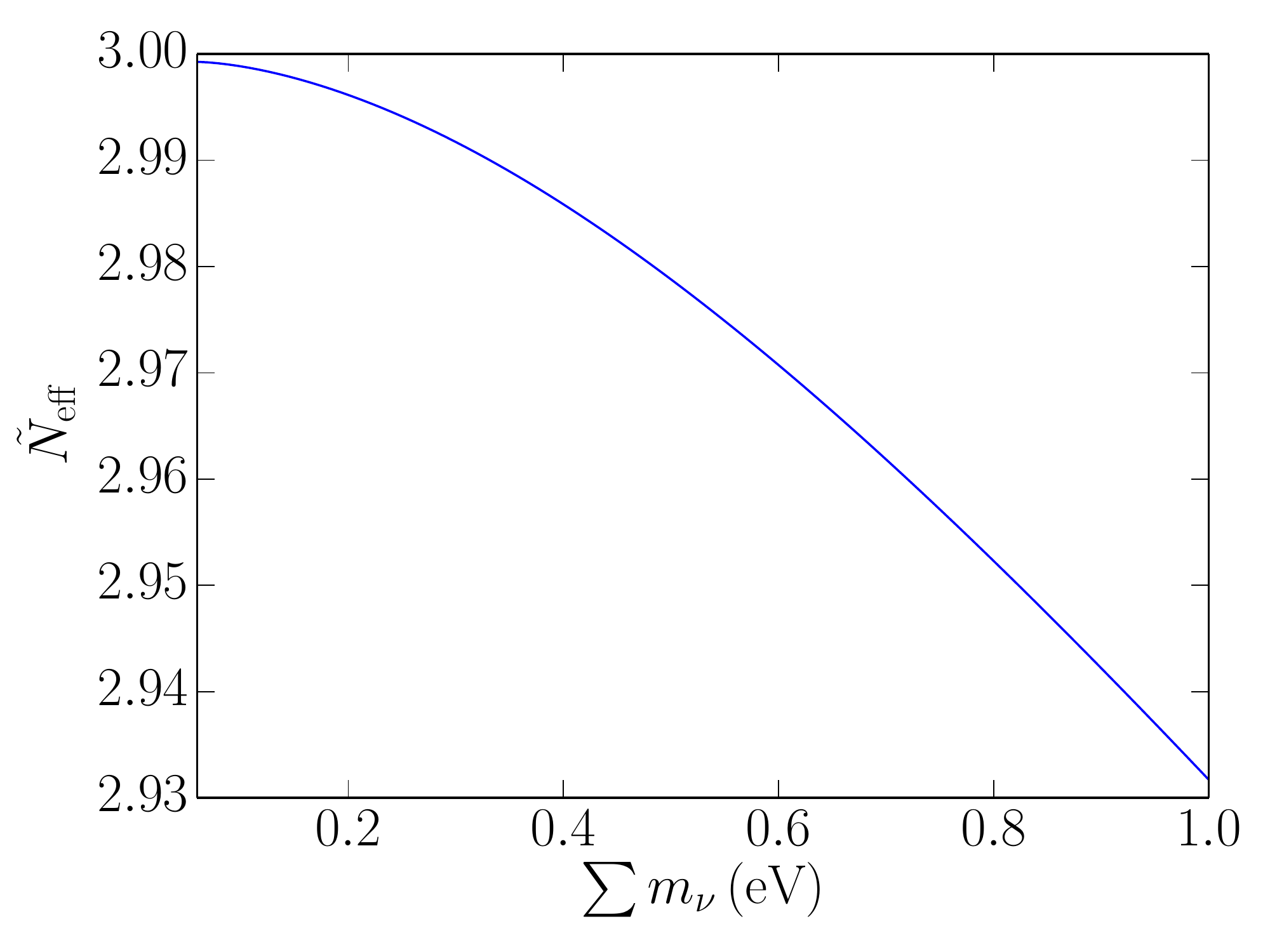}
   \caption{\label{fig:tneffvsummnu} \tneff as a function of
   \summnu.  The minimum value for \summnu is $\sim0.06\,{\rm eV}$
   for the normal mass hierarchy.}
\end{figure}

\begin{figure}[b]
\includegraphics[width=0.5\textwidth]{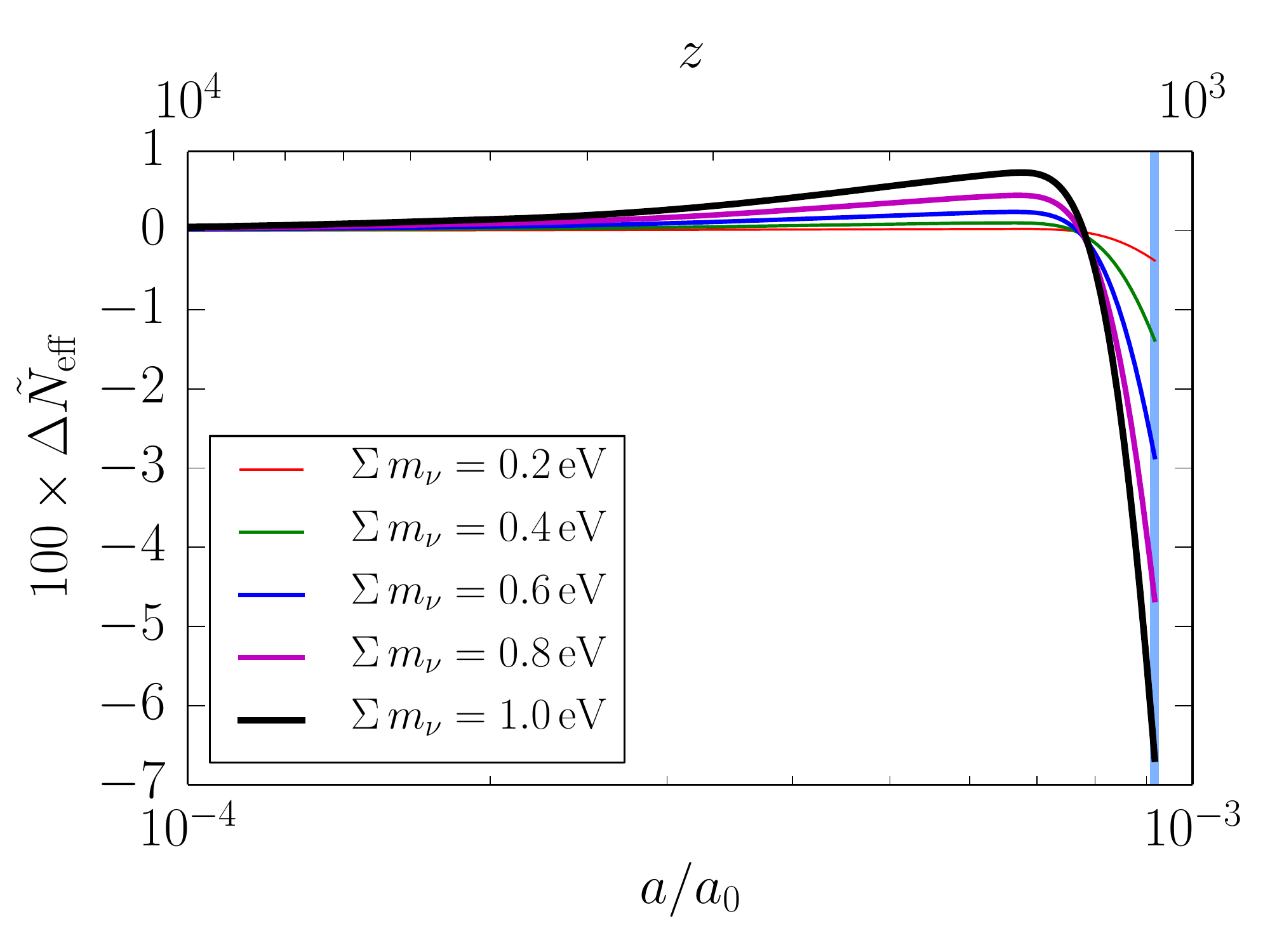}
\caption{\label{fig:neff_vs_a_non}(Color online.) The change in
   \neffobs, $\Delta\neffobs$, is given as a function of scale factor
   ratio, $a/a_0$, and redshift, $z$, (at top).  The primordial helium
   mass fraction and vertical bar are identical to Fig.\
   \ref{fig:xe_vs_a_0}.  For each value of \summnu, $\Delta\neffobs$
   is initially positive.  $\Delta\neffobs$ becomes negative once the
   recombination histories of Fig.\ \ref{fig:xe_vs_a_non} differ from
the massless case.}
\end{figure}

\begin{figure}[b]
\includegraphics[width=0.5\textwidth]{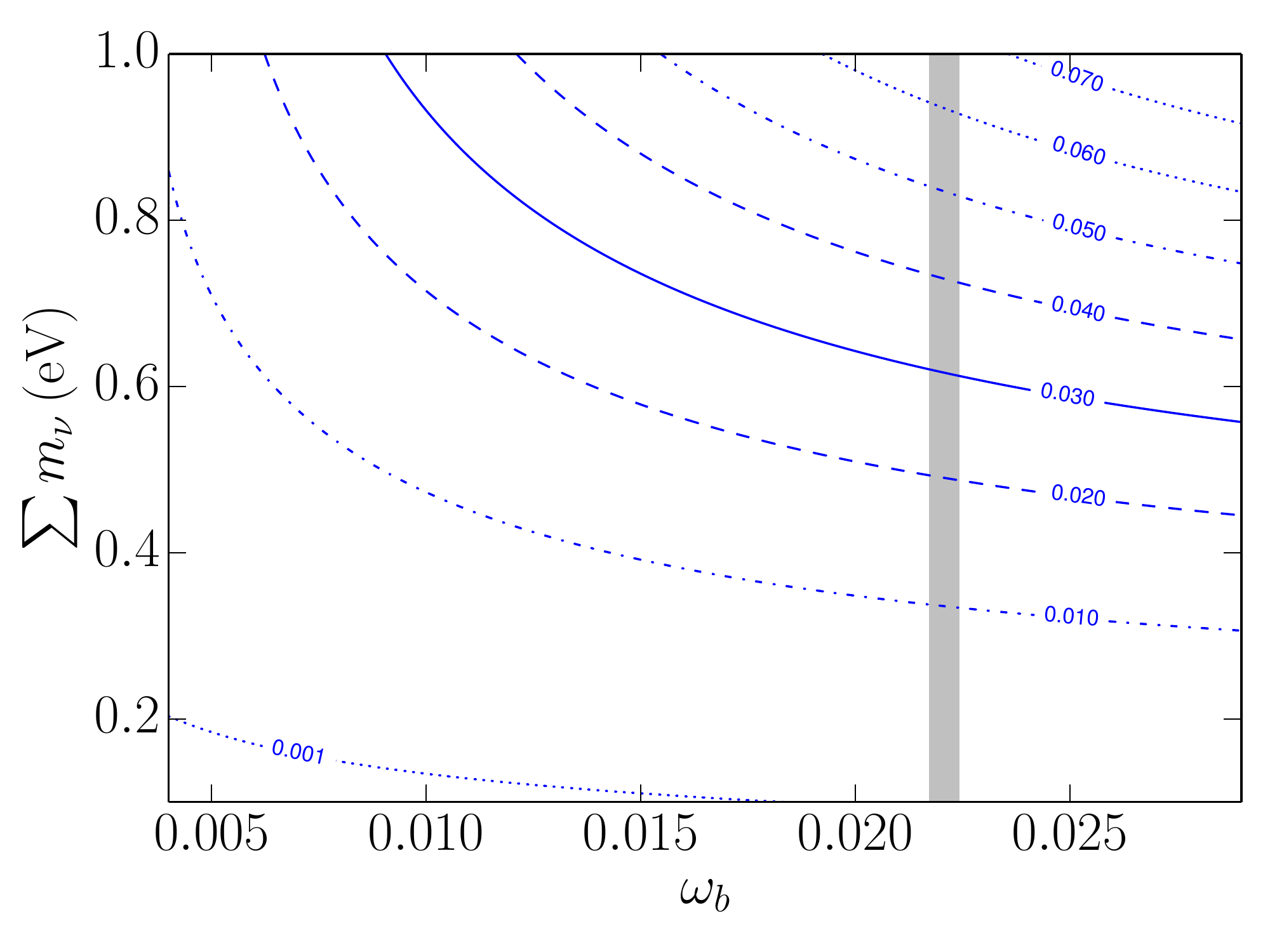}
\caption{\label{fig:contour_mass}(Color online.) Contours of constant
   $-\Delta\neffobs$ in the \summnu vs.\ \barnum parameter space.  The
   shaded, vertical bar corresponds to the $1\sigma$ error for \barnum
   \cite{2013arXiv1303.5076P}.  
}
\end{figure}

Using Eq.\ \eqref{eqn:rho_nu} with neutrino masses taken as described
above for a given value of \summnu and the known relations for
$\rho_\gamma$ and $\rho_b$ we compute the quantities $r_s$ and $r_d$
from Eqs.\ \eqref{eqn:rs} and \eqref{eqn:rd}, respectively.  We
determine the inferred measure of $\neff$, \neffobs, by choosing
\neffth in Eq.\ \eqref{eqn:baby_formula} to reproduce this ratio of
$r_s/r_d$, which is a monotonically decreasing, invertible function of
\neffth; see Fig.\ \ref{fig:rsrdvneffth}. The quantity \neffobs reduces
to \neffth for massless, decoupled neutrinos. There is clearly no
requirement that $\neffobs$ be independent of scale factor $a$.  It
has been motivated here by the need to characterize massive neutrinos
but it is also applicable to non-standard cosmologies with
constituents that may be far from equilibrium.  Using the ratio
$r_s/r_d$ in the determination of \neffobs avoids any reference to the
angular diameter distance to last scattering and, therefore,
dependence on the dark energy equation of
state\cite{2013PhRvD..87h3008H}. In contrast to
Ref.\ \cite{2013PhRvD..87h3008H} we do not change the value of the
primordial helium abundance, $Y_P$ to keep $\theta_d$ fixed.  This
fact and $\Delta (r_s/r_d)\sim -\Delta\neffobs$ (since $r_s/r_d$ is
monotonically decreasing with \neffobs) means that a larger Hubble
rate would imply a larger value of \neffobs. In fact, as suggested
in Fig.\ \ref{fig:xe_vs_a_non}, the dependence of $n_e(a)$ on the
increased energy density from the neutrino rest mass dominates the
explicit dependence on $H(a)$ in Eq.\ \eqref{eqn:rd}. This has the
consequence that a larger Hubble rate results in a {\em larger} ratio of
$r_s/r_d$; see
Figs.\ \ref{fig:tneffvsummnu} to \ref{fig:contour_mass} and
discussion below.

In order to investigate physics beyond the standard models of particle
physics and cosmology, we have formulated a self-consistent approach
that is not constrained to minimal extensions to the standard model.
To do so, we simulate the early universe from weak decoupling through
Big Bang Nucleosynthesis (BBN) to photon decoupling using the {\tt
BURST} code\cite{GFKP-5pts:2014mn}.  This treatment self-consistently
incorporates binned, general, momentum occupation probabilities for
each of six neutrino species ($\nu_e$, $\bar{\nu}_e$, $\nu_\mu$,
$\bar{\nu}_\mu$, $\nu_\tau$, and $\bar{\nu}_\tau$) and a Boltzmann
treatment of neutrino scattering, absorption and emission processes to
evolve the early universe.  

In this treatment, neutrinos decouple from the $\gamma$, $e^\pm$ plasma at high
temperatures, $1\!\lesssim\! T\!  \lesssim\! 3$ MeV, with ultra-relativistic
kinematics\cite{2009PhRvL.102t1303F, 1990bmc..book.....D} as expected. The
computation of the helium abundance $Y_P$ in this treatment
however is more nuanced than in the standard cosmology.  Here $Y_P$ is not
simply a function of $\neffth$ and $\omega_b$.  Assuming zero lepton numbers
and an adopted world-average neutron lifetime of $886$ s, our calculations give
a $^4$He primordial mass fraction $Y_P=0.242$ taking the baryon number
$\Omega_bh^2\equiv\barnum = 0.022068$ from the Ref.\ \cite{2013arXiv1303.5076P}
best-fit. This is consistent with the observationally inferred primordial
helium abundance \cite{Izotov:2010ns,2013JCAP...11..017A}. Although we take the
neutrinos to decouple in weak eigenstates, i.e.\ flavor states, we write their
occupation probabilities in the mass eigenbasis. Since we are assuming the
neutrinos have identical thermal spectra with zero-chemical potential in the
weak eigenstates, we can use the same occupation probabilities for the mass
eigenstates at a given momentum
$p$\cite{2002NuPhB.632..363D,2009PhRvL.102t1303F}.  

Before considering the effect of the neutrino mass on \neffobs, the
deduced radiation energy density, we estimate its effect on \neffth.
The cosmological constraint (at the level of $2\sigma$) on the sum of
the light neutrino masses is $\summnu\le0.23$ eV
\cite{2013arXiv1303.5076P}.  If we take \summnu to be at this upper
limit and assume degenerate mass eigenvalues, each neutrino has an
associated mass $\sim0.08\,{\rm eV}$.  We see that the neutrino rest
masses and temperatures at photon decoupling
($T_\gamma\approx0.2\,{\rm eV}$, $T_\nu\approx0.15\,{\rm eV}$) are
coincidentally at the same scale, meaning that neutrinos can not be
treated either as pure matter or pure radiation.  An individual
neutrino has an average momentum of $\sim0.5\,{\rm eV}$ at photon
decoupling.  As a consequence, we expect fractional corrections to the
relativistic neutrino energy density stemming from neutrino rest mass
to be $\sim m^2/2p^2\sim 0.01$, with a concomitant change to \neffth
of $\sim3\times0.01\sim +0.03$.  If we were to unphysically classify
the entire neutrino energy density into $\rho_\mathrm{rad}$ the
corresponding change to \neffth would be $\Delta\neffth\equiv\neffth -
3\simeq\frac{5}{7\pi^2} \left(\frac{11}{4}\right)^{2/3}
\sum_{i=1}^3\left(\frac{m_i}{T_\gamma}\right)^2$.  We arrive then at
$\Delta\neffth\simeq 0.04$ for $\summnu = 0.23\,{\rm eV}$, a change
consistent with the simple kinematic estimate above, and not to be
confused with $\Delta\neffth\approx0.046$ stemming from
non-equilibrium neutrino scattering and quantum-electrodynamics effects
inherent in
Refs.\ \cite{1999NuPhB.543..269D,1982NuPhB.209..372C,1999PhRvD..59j3502L}.

Figure \ref{fig:tneffvsummnu} shows \tneff versus \summnu.  Contrary to the
expectation of increased \tneff in our previous scaling and estimate arguments,
we observe a monotonic decrease in \tneff with increasing \summnu.  We denote
this phenomenon the neutrino mass/recombination (\numr) effect.  For
illustrative purposes in Fig.\ \ref{fig:neff_vs_a_non}, we treat $\Delta\tneff$
as a quantity to be determined at any epoch, although \neff is only observed at
photon decoupling.  The neutrino rest mass has no discernible effect on
\neffobs at early epochs, at small $a/a_0$.  At larger values of
$a/a_0\,(\sim5\times10^{-4})$, the extra energy density from the neutrino rest
masses produces a larger \neffobs in accordance with Eq.\ \eqref{eqn:scaling}.
If we were to extrapolate this
evolution trend to the epoch of photon decoupling, we would find a value of
$\Delta\neffobs > 0$.  The \numr effect intervenes to modify this extrapolation
and results in $\Delta\neffobs<0$ at $a_{\gamma d}$.

Each evolution curve for $\Delta\neffobs$ in Fig.\
\ref{fig:neff_vs_a_non} corresponds to an evolution curve for $X_e(a)$
in Fig.\ \ref{fig:xe_vs_a_non} for various values of $\summnu$.  The
smallest value of \summnu produces the smallest change in $X_e$, which
subsequently changes $\Delta\neffobs$ the least.  Conversely, the
largest value of \summnu produces the largest change in $X_e$, which
changes $\Delta\neffobs$ the most.  From the curves in Fig.\
\ref{fig:neff_vs_a_non}, it is clear that the effect of neutrino rest
mass in producing a higher $X_e$ at freeze-out overwhelms the effect
of the extra energy density, thereby decreasing \neffobs at photon
decoupling, i.e.\ at $a = a_{\gamma d}$, the vertical bar in Figs.\
\ref{fig:xe_vs_a_non} and \ref{fig:neff_vs_a_non}.

There are several interesting features to note in Fig.\
\ref{fig:neff_vs_a_non}.  Each curve in Fig.\ \ref{fig:neff_vs_a_non} goes
through $\Delta\neffobs = 0$ near the value
$a/a_0\sim(7.65\pm0.10)\times10^{-4}$.  The larger the value of \summnu, the
higher the curvature of the function $\Delta\neffobs(a)$.  For values of
$a/a_0$ above which $\Delta\neffobs = 0$, the slope of $\Delta\neffobs(a)$ is a
rapidly decreasing function of \summnu.
%{\color{red}  The thick black curve, corresponding to $\summnu=1.0\,{\rm eV}$,
%has the most extreme zenith and nadir, and therefore undergoes the largest
%correction for any mass value in Fig.\ \ref{fig:neff_vs_a_non}.  The \numr
%correction becomes less pronounced with decreasing \summnu for the range of
%mass values we explore.}
We note that for $\summnu = 0.23$ eV, the preferred upper limit from
Ref.\ \cite{2013arXiv1303.5076P}, we find $\Delta\neffobs = -0.005$. This effect
is certainly below present sensitivities of CMB observations.
%but we note the following important points.
Next-generation CMB measurements, however, aspire to percent level accuracy in
determinations of the relativistic energy density\cite{Abazajian201566}. The
exquisite sensitivity of the \numr effect on $\Delta\neffobs(a_{\gamma d})$
suggests that it may be an important component in future precision
determinations of cosmological parameters.

%We should clarify a potential point of confusion that might cause one
%to conclude that \neffobs would decrease with increasing \summnu. The
%false argument to which we refer is that neutrinos with rest mass
%$\mathcal{O}(0.1\text{ eV})$, entering a non-relativistic kinematical
%regime at photon decoupling would lower \neffth as calculated in
%Eq.\eqref{eqn:baby_formula}. Though this line of reasoning would give
%the right sign for $\Delta\neffobs$, it is incorrect since neither $r_s$
%nor $r_d$ depend explicitly on $\rho_m$, the matter component of the
%energy density.

%We clarify a potential point of confusion that might cause one
%to conclude that \neffobs would decrease with increasing \summnu. 
%Neutrinos with rest mass
%$\mathcal{O}(0.1\text{ eV})$ are entering the non-relativistic kinematical
%regime at photon decoupling.  One would think that
%the transition from radiation to matter
%decreases the number of degrees of freedom, and
%subsequently the energy density.
%We expand Eq.\ref{eqn:rho_nu}
%in terms of $(p/m_\nu)^2$:
%\begin{align}
%\rho(m_\nu\ne0)
%   &\simeq\frac{1}{2\pi^2} \int_0^\infty dp\,p^2
%      \frac{1}{e^{p/T_\nu} + 1}p^2
%      \left(1+\frac{m_\nu^2}{2p^2}\right)\\
%   &=\rho(m_\nu=0) + \mathcal{O}(m_\nu^2)
%\end{align}
%The expansion shows that there is no deprecation in the
%ultra-relativistic contribution to $\rho$.
%There is no negative change in \neffth from the
%radiation-matter transition of neutrinos.  The negative
%change in \neffobs is due to the different recombination
%history displayed in Fig.\ \ref{fig:xe_vs_a_non}.

As mentioned earlier, we do not constrain $a_{\gamma d}$
to maintain a uniform optical depth $\tau(a_{\gamma d})$
\begin{align}
   \label{eqn:opt_depth} 
   \tau(a_{\gamma d}) = \int_{a_{\gamma d}}^{a_0} \frac{da}{a^2H}
   a\,n_e(a)\,\sigma_T
   \equiv 1,
\end{align} 
when comparing different values for \summnu.
%Note that we are neglecting reionization effects on $n_e(a)$.
Note that this definition of $\tau(a_{\gamma d})$
does not include reionization effects on $n_e(a)$.
We should emphasize that
each curve in Figs.\ \ref{fig:xe_vs_a_non} and \ref{fig:neff_vs_a_non}
is calculated using the same value for the scale factor of last
scattering $a_{\gamma d} = 9.162 \times 10^{-4}$ (corresponding to
$z=1090.43$).  This is not self consistent, strictly speaking, but we
have verified that the effect on \neffobs, due to the differences in
$n_e$ and $H$, is negligible.  If we impose the constraint in
Eq.\ \eqref{eqn:opt_depth}, we find $a_{\gamma d}$ decreases by a few
parts in $10^4$ for $\summnu=0.23$ eV, which has an insignificant
effect on \neffobs.

Up to this point in the present analysis we have not considered
variation of the primordial helium mass fraction $Y_P$ since BBN
occurs at high enough temperatures that the neutrinos are effectively
massless. If we consider, however, cosmological parameters that affect
$Y_P$ we can examine the dependence of ionization freeze-out (and
subsequent alteration of \neffobs) on both \summnu and $Y_P$
simultaneously. A direct way to vary $Y_P$ is to consider changes to
the baryon number \barnum. In the range of values of \barnum that
we're interested in, $Y_P$ is a monotonically increasing function of
\barnum.

Figure \ref{fig:contour_mass} shows a contour plot of $\Delta\neffobs$
in the \summnu versus \barnum parameter space; contours correspond to
constant values of $-\Delta\neffobs$.  Varying \barnum requires new
computations of $Y_P$ from BBN and $X_e(a)$ from recombination.
Changing \summnu requires a new computation of $X_e$ but no new
computation of $Y_P$.  As a consequence, we compute BBN with the {\tt
BURST} code only once for a given \barnum, and compute the
recombination history for each pair (\barnum, \summnu). Holding
\summnu fixed, the change in $|\Delta\neffobs|$ increases with
increasing \barnum due to the different recombination histories
effecting a change in $r_d$. Note that the
change in $r_s$
does not completely compensate for
the change in $r_d$.  The shaded vertical region in Fig.\
\ref{fig:contour_mass} is the $1\sigma$ range of \barnum given by
Ref.\ \cite{2013arXiv1303.5076P}, but we explore a larger range in the
\barnum parameter space to illustrate the dependence of \neffobs on
\summnu and \barnum.  An interesting feature of these curves is their
increasing curvature with decreasing \barnum and increasing \summnu.
This is a consequence of an enhancement of the \numr effect with
increasing \barnum: as \summnu increases, the change in \neffobs is
faster for higher values of \barnum.  
%This is perhaps a surprising result given the form of
%Eq.\eqref{eqn:rd} for $r_d$.  A scaling of $Y_P$ would appear to lead
%to an increase in $r_d$ since $n_e \propto (1-\tfrac{1}{2}Y_P)$.
%This would, of course, lead to a smaller ratio of $r_s/r_d$ and
%therefore a larger \neffobs, in contradiction to what is revealed in
%Fig.\ \ref{fig:contour_mass}. Numerical computations using our
%recombination history model show, however, that $n_e(a_{\gamma d})$
%is a decreasing function of \barnum (or $Y_P$) in the range of
%\barnum considered here.  This explains the \numr enhancement of
%Fig.\ \ref{fig:contour_mass}.

%We might ask whether the change in $\Delta\neffobs$ follows the change
%in $Y_P$. In fact, we notice an amplification of the increase in
%$|\Delta\neff|$ with increasing \barnum. $Y_P$ increases with
%increasing \barnum, thereby decreasing the number density of free
%electrons and increasing the diffusion length.  As compared to the
%massless case, the {\it relative} change in the diffision length for
%massive neutrinos is agnostic to \barnum, but the {\it absolute}
%change in the sound horizon is agnostic to \barnum to the order
%considered here.  Hence, there is a larger change in the ratio
%$r_s/r_d$ for larger \barnum at a given \summnu which produces the
%larger $|\Delta\neff|$.

We have discussed two ways in which neutrino rest mass affects
measurable quantities at photon decoupling.  First, neutrino rest mass
drives an earlier recombination freeze-out resulting in a higher
free--electron fraction.  Second, this effect is enhanced with
increasing \barnum stemming from a self-consistently calculated
recombination history.  As radiation energy density is not a directly
measurable quantity, we use observable quantities to indirectly arrive
at the radiation energy density.  For this purpose, we choose the
ratio of the sound horizon to the photon diffusion length. Photon
diffusion is sensitive to the recombination history,
which requires a Boltzmann-equation treatment.  
We find a non-trivial evolution of $\Delta\neffobs$ with scale factor,
as shown in Fig.\ \ref{fig:neff_vs_a_non}.  Note that the evolution of
\neffobs shown in this figure does not reflect a {\em kinematical}
evolution of the radiation energy density with a massive component.
The trends evidenced in this figure are a consequence of the \numr
effect.

Self-consistency is a primary motivation for defining the radiation
energy density parameter \neffobs in terms of the ratio $r_s/r_d$; it
generalizes the \neffth parameter to the massive neutrino case.
Further, \neffobs is defined for general energy densities and
non-equilibrium distribution functions where the neutrino temperature
is undefined. Moreover, \neffobs makes no
assumption regarding the underlying cosmological model.
We use \neffobs to relate the sound horizon and photon
diffusion length to predictions made by the standard cosmological
model via the parameter \neffth.

The \numr effect is an example of a recurring phenomenon in cosmology:
an increase in the expansion rate leads to an earlier epoch of
freeze-out. This effect was revealed in the present context by using
\neffobs to {\em infer} the cosmic radiation energy content from
observable CMB data, rather than treating \neffth as an {\em input}.
The procedure we describe here differs from that adopted by the public
Boltzmann codes.  CAMB\cite{Howlett:2012mh}, for example, includes
options to evolve massive neutrino energy density through the epoch of
recombination and requires \neffth to be provided as an input.

Depending on \summnu, the concomitant changes in
ionization equilibrium and \neffobs discussed here may be
within the sensitivity of the next generation CMB experiments when
polarization effects are taken into account\cite{2013arXiv1303.5076P,
2014ApJ...794..171T,SPT:2003IA,Abazajian201566}.  CMB
precision is planned to be increased to the $\neffobs\sim1\%$ level
which would probe both massive active neutrinos and other possible
components of dark radiation. Scenarios with sterile neutrinos and
other very weakly coupled light massive species with masses larger
than those associated with the active neutrinos could enhance the
effects discussed here.  However, depending on their masses and their
flavor mixing with active species, sterile neutrinos could have number
densities and energy spectra which differ from those of active
neutrinos \cite{2005PhRvD..72f3004A, 2006PhRvL..97n1301K,
2010PhRvL.105r1301H, 2011JCAP...09..034H, 2014PhRvD..90h3503D,
2005NuPhB.708..215C}, complicating the analysis given here.

%{\color{red} Adding rest mass to neutrinos changes the epoch of
%   matter-radiation equality.  If $z_{\rm eq}$ is the redshift of
%   matter radiation equality, then when $z=z_{\rm eq}$:
%   \begin{equation}
%      \rho_\gamma+\rho_\nu^{(0)}=\rho_b+\rho_c+\rho_\nu^{(1)},
%   \end{equation} where $\rho_\nu^{(0)}$ is the ultra-relativistic
%   component and $\rho_\nu^{(1)}$ is the first-order correction for
%   nonzero rest mass of the neutrino energy density.  Our estimates
%   place $z_{\rm eq}=3420.54$ for massless neutrinos with closure
%   fractions $\Omega_bh^2=0.022068$ and $\Omega_ch^2=0.12029$.  When
%   $\summnu=0.23 \text{ eV}$, we find only a small change in $z_{\rm
%   eq}$ (relative difference $\lesssim6\times10^{-4}$).  We do not
%consider neutrino-rest-mass effects on $z_{\rm eq}$ in this letter.}

We would like to acknowledge the Institutional Computing Program at
Los Alamos National Laboratory for use of their HPC cluster resources.
This work was supported in part by NSF grant PHY-1307372 at UC San
Diego, by the Los Alamos National Laboratory Institute for Geophysics,
Space Sciences and Signatures subcontract 257842, and the National
Nuclear Security Administration of the U.S.\ Department of Energy at
Los Alamos National Laboratory under Contract No.\ DE-AC52-06NA25396.
We thank J.J.\ Cherry, Amit Yadav, and Lloyd Knox for helpful
discussions.  We would also like to thank the anonymous referees for
their useful comments.
    
%\nocite{*}

\bibliography{neff_summnu_bib}% Produces the bibliography via BibTeX.

\begin{thebibliography}{32}
\expandafter\ifx\csname natexlab\endcsname\relax\def\natexlab#1{#1}\fi
\expandafter\ifx\csname bibnamefont\endcsname\relax
  \def\bibnamefont#1{#1}\fi
\expandafter\ifx\csname bibfnamefont\endcsname\relax
  \def\bibfnamefont#1{#1}\fi
\expandafter\ifx\csname citenamefont\endcsname\relax
  \def\citenamefont#1{#1}\fi
\expandafter\ifx\csname url\endcsname\relax
  \def\url#1{\texttt{#1}}\fi
\expandafter\ifx\csname urlprefix\endcsname\relax\def\urlprefix{URL }\fi
\providecommand{\bibinfo}[2]{#2}
\providecommand{\eprint}[2][]{\url{#2}}

\bibitem[{\citenamefont{{Dolgov} et~al.}(1999)\citenamefont{{Dolgov}, {Hansen},
  and {Semikoz}}}]{1999NuPhB.543..269D}
\bibinfo{author}{\bibfnamefont{A.~D.} \bibnamefont{{Dolgov}}},
  \bibinfo{author}{\bibfnamefont{S.~H.} \bibnamefont{{Hansen}}},
  \bibnamefont{and} \bibinfo{author}{\bibfnamefont{D.~V.}
  \bibnamefont{{Semikoz}}}, \bibinfo{journal}{Nucl. Phys. B}
  \textbf{\bibinfo{volume}{543}}, \bibinfo{pages}{269} (\bibinfo{year}{1999}),
  \eprint{hep-ph/9805467}.

\bibitem[{\citenamefont{{Cambier} et~al.}(1982)\citenamefont{{Cambier},
  {Primack}, and {Sher}}}]{1982NuPhB.209..372C}
\bibinfo{author}{\bibfnamefont{J.-L.} \bibnamefont{{Cambier}}},
  \bibinfo{author}{\bibfnamefont{J.~R.} \bibnamefont{{Primack}}},
  \bibnamefont{and} \bibinfo{author}{\bibfnamefont{M.}~\bibnamefont{{Sher}}},
  \bibinfo{journal}{Nucl. Phys. B} \textbf{\bibinfo{volume}{209}},
  \bibinfo{pages}{372} (\bibinfo{year}{1982}).

\bibitem[{\citenamefont{{Lopez} and {Turner}}(1999)}]{1999PhRvD..59j3502L}
\bibinfo{author}{\bibfnamefont{R.~E.} \bibnamefont{{Lopez}}} \bibnamefont{and}
  \bibinfo{author}{\bibfnamefont{M.~S.} \bibnamefont{{Turner}}},
  \bibinfo{journal}{\prd} \textbf{\bibinfo{volume}{59}}, \bibinfo{eid}{103502}
  (\bibinfo{year}{1999}), \eprint{astro-ph/9807279}.

\bibitem[{\citenamefont{{Mangano} et~al.}(2002)\citenamefont{{Mangano},
  {Miele}, {Pastor}, and {Peloso}}}]{Mangano:3.040}
\bibinfo{author}{\bibfnamefont{G.}~\bibnamefont{{Mangano}}},
  \bibinfo{author}{\bibfnamefont{G.}~\bibnamefont{{Miele}}},
  \bibinfo{author}{\bibfnamefont{S.}~\bibnamefont{{Pastor}}}, \bibnamefont{and}
  \bibinfo{author}{\bibfnamefont{M.}~\bibnamefont{{Peloso}}},
  \bibinfo{journal}{PLB} \textbf{\bibinfo{volume}{534}}, \bibinfo{pages}{8}
  (\bibinfo{year}{2002}), \eprint{astro-ph/0111408}.

\bibitem[{\citenamefont{{Shimon} et~al.}(2010)\citenamefont{{Shimon}, {Miller},
  {Kishimoto}, {Smith}, {Fuller}, and {Keating}}}]{2010JCAP...05..037S}
\bibinfo{author}{\bibfnamefont{M.}~\bibnamefont{{Shimon}}},
  \bibinfo{author}{\bibfnamefont{N.~J.} \bibnamefont{{Miller}}},
  \bibinfo{author}{\bibfnamefont{C.~T.} \bibnamefont{{Kishimoto}}},
  \bibinfo{author}{\bibfnamefont{C.~J.} \bibnamefont{{Smith}}},
  \bibinfo{author}{\bibfnamefont{G.~M.} \bibnamefont{{Fuller}}},
  \bibnamefont{and} \bibinfo{author}{\bibfnamefont{B.~G.}
  \bibnamefont{{Keating}}}, \bibinfo{journal}{\jcap}
  \textbf{\bibinfo{volume}{5}}, \bibinfo{eid}{037} (\bibinfo{year}{2010}),
  \eprint{1001.5088}.

\bibitem[{\citenamefont{{Hou} et~al.}(2013)\citenamefont{{Hou}, {Keisler},
  {Knox}, {Millea}, and {Reichardt}}}]{2013PhRvD..87h3008H}
\bibinfo{author}{\bibfnamefont{Z.}~\bibnamefont{{Hou}}},
  \bibinfo{author}{\bibfnamefont{R.}~\bibnamefont{{Keisler}}},
  \bibinfo{author}{\bibfnamefont{L.}~\bibnamefont{{Knox}}},
  \bibinfo{author}{\bibfnamefont{M.}~\bibnamefont{{Millea}}}, \bibnamefont{and}
  \bibinfo{author}{\bibfnamefont{C.}~\bibnamefont{{Reichardt}}},
  \bibinfo{journal}{\prd} \textbf{\bibinfo{volume}{87}}, \bibinfo{eid}{083008}
  (\bibinfo{year}{2013}), \eprint{1104.2333}.

\bibitem[{\citenamefont{Birrell et~al.}(2014)\citenamefont{Birrell, Yang, Chen,
  and Rafelski}}]{Birrell.PhysRevD.89.023008}
\bibinfo{author}{\bibfnamefont{J.}~\bibnamefont{Birrell}},
  \bibinfo{author}{\bibfnamefont{C.-T.} \bibnamefont{Yang}},
  \bibinfo{author}{\bibfnamefont{P.}~\bibnamefont{Chen}}, \bibnamefont{and}
  \bibinfo{author}{\bibfnamefont{J.}~\bibnamefont{Rafelski}},
  \bibinfo{journal}{Phys. Rev. D} \textbf{\bibinfo{volume}{89}},
  \bibinfo{pages}{023008} (\bibinfo{year}{2014}).

\bibitem[{\citenamefont{{Hou} et~al.}(2014)\citenamefont{{Hou}, {Reichardt},
  {Story}, {Follin}, {Keisler}, {Aird}, {Benson}, {Bleem}, {Carlstrom}, {Chang}
  et~al.}}]{2014ApJ...782...74H}
\bibinfo{author}{\bibfnamefont{Z.}~\bibnamefont{{Hou}}},
  \bibinfo{author}{\bibfnamefont{C.~L.} \bibnamefont{{Reichardt}}},
  \bibinfo{author}{\bibfnamefont{K.~T.} \bibnamefont{{Story}}},
  \bibinfo{author}{\bibfnamefont{B.}~\bibnamefont{{Follin}}},
  \bibinfo{author}{\bibfnamefont{R.}~\bibnamefont{{Keisler}}},
  \bibinfo{author}{\bibfnamefont{K.~A.} \bibnamefont{{Aird}}},
  \bibinfo{author}{\bibfnamefont{B.~A.} \bibnamefont{{Benson}}},
  \bibinfo{author}{\bibfnamefont{L.~E.} \bibnamefont{{Bleem}}},
  \bibinfo{author}{\bibfnamefont{J.~E.} \bibnamefont{{Carlstrom}}},
  \bibinfo{author}{\bibfnamefont{C.~L.} \bibnamefont{{Chang}}},
  \bibnamefont{et~al.}, \bibinfo{journal}{\apj} \textbf{\bibinfo{volume}{782}},
  \bibinfo{eid}{74} (\bibinfo{year}{2014}), \eprint{1212.6267}.

\bibitem[{\citenamefont{Howlett et~al.}(2012)\citenamefont{Howlett, Lewis,
  Hall, and Challinor}}]{Howlett:2012mh}
\bibinfo{author}{\bibfnamefont{C.}~\bibnamefont{Howlett}},
  \bibinfo{author}{\bibfnamefont{A.}~\bibnamefont{Lewis}},
  \bibinfo{author}{\bibfnamefont{A.}~\bibnamefont{Hall}}, \bibnamefont{and}
  \bibinfo{author}{\bibfnamefont{A.}~\bibnamefont{Challinor}},
  \bibinfo{journal}{JCAP} \textbf{\bibinfo{volume}{1204}}, \bibinfo{pages}{027}
  (\bibinfo{year}{2012}), \eprint{1201.3654}.

\bibitem[{\citenamefont{{Lesgourgues} and {Pastor}}(2012)}]{Lesgourges:2012nm}
\bibinfo{author}{\bibfnamefont{J.}~\bibnamefont{{Lesgourgues}}}
  \bibnamefont{and} \bibinfo{author}{\bibfnamefont{S.}~\bibnamefont{{Pastor}}},
  \bibinfo{journal}{Adv.\ High En.\ Phys.} \textbf{\bibinfo{volume}{2012}},
  \bibinfo{pages}{608515} (\bibinfo{year}{2012}), \eprint{1212.6154}.

\bibitem[{\citenamefont{{Planck Collaboration}}(2014)}]{2013arXiv1303.5076P}
\bibinfo{author}{\bibnamefont{{Planck Collaboration}}}, \bibinfo{journal}{\aap}
  \textbf{\bibinfo{volume}{571}}, \bibinfo{eid}{A16} (\bibinfo{year}{2014}),
  \eprint{1303.5076}.

\bibitem[{\citenamefont{{Hu} and {Sugiyama}}(1996)}]{Hu.1996ApJ.471.542H}
\bibinfo{author}{\bibfnamefont{W.}~\bibnamefont{{Hu}}} \bibnamefont{and}
  \bibinfo{author}{\bibfnamefont{N.}~\bibnamefont{{Sugiyama}}},
  \bibinfo{journal}{\apj} \textbf{\bibinfo{volume}{471}}, \bibinfo{pages}{542}
  (\bibinfo{year}{1996}), \eprint{astro-ph/9510117}.

\bibitem[{\citenamefont{{Peebles}}(1968)}]{1968ApJ...153....1P}
\bibinfo{author}{\bibfnamefont{P.~J.~E.} \bibnamefont{{Peebles}}},
  \bibinfo{journal}{\apj} \textbf{\bibinfo{volume}{153}}, \bibinfo{pages}{1}
  (\bibinfo{year}{1968}).

\bibitem[{\citenamefont{{Zeldovich} et~al.}(1968)\citenamefont{{Zeldovich},
  {Kurt}, and {Syunyaev}}}]{1968ZhETF..55..278Z}
\bibinfo{author}{\bibfnamefont{Y.~B.} \bibnamefont{{Zeldovich}}},
  \bibinfo{author}{\bibfnamefont{V.~G.} \bibnamefont{{Kurt}}},
  \bibnamefont{and} \bibinfo{author}{\bibfnamefont{R.~A.}
  \bibnamefont{{Syunyaev}}}, \bibinfo{journal}{Zhurnal Eksperimentalnoi i
  Teoreticheskoi Fiziki} \textbf{\bibinfo{volume}{55}}, \bibinfo{pages}{278}
  (\bibinfo{year}{1968}).

\bibitem[{\citenamefont{{Seager} et~al.}(1999)\citenamefont{{Seager},
  {Sasselov}, and {Scott}}}]{1999ApJ...523L...1S}
\bibinfo{author}{\bibfnamefont{S.}~\bibnamefont{{Seager}}},
  \bibinfo{author}{\bibfnamefont{D.~D.} \bibnamefont{{Sasselov}}},
  \bibnamefont{and} \bibinfo{author}{\bibfnamefont{D.}~\bibnamefont{{Scott}}},
  \bibinfo{journal}{\apjl} \textbf{\bibinfo{volume}{523}}, \bibinfo{pages}{L1}
  (\bibinfo{year}{1999}), \eprint{astro-ph/9909275}.

\bibitem[{\citenamefont{Zahn and Zaldarriaga}(2003)}]{Zahn.PhysRevD.67.063002}
\bibinfo{author}{\bibfnamefont{O.}~\bibnamefont{Zahn}} \bibnamefont{and}
  \bibinfo{author}{\bibfnamefont{M.}~\bibnamefont{Zaldarriaga}},
  \bibinfo{journal}{Phys. Rev. D} \textbf{\bibinfo{volume}{67}},
  \bibinfo{pages}{063002} (\bibinfo{year}{2003}).

\bibitem[{\citenamefont{{Mangano} et~al.}(2005)\citenamefont{{Mangano},
  {Miele}, {Pastor}, {Pinto}, {Pisanti}, and {Serpico}}}]{neff:3.046}
\bibinfo{author}{\bibfnamefont{G.}~\bibnamefont{{Mangano}}},
  \bibinfo{author}{\bibfnamefont{G.}~\bibnamefont{{Miele}}},
  \bibinfo{author}{\bibfnamefont{S.}~\bibnamefont{{Pastor}}},
  \bibinfo{author}{\bibfnamefont{T.}~\bibnamefont{{Pinto}}},
  \bibinfo{author}{\bibfnamefont{O.}~\bibnamefont{{Pisanti}}},
  \bibnamefont{and} \bibinfo{author}{\bibfnamefont{P.~D.}
  \bibnamefont{{Serpico}}}, \bibinfo{journal}{Nuclear Physics B}
  \textbf{\bibinfo{volume}{729}}, \bibinfo{pages}{221} (\bibinfo{year}{2005}),
  \eprint{hep-ph/0506164}.

\bibitem[{\citenamefont{{Grohs} et~al.}(2015)\citenamefont{{Grohs}, {Fuller},
  {Kishimoto}, and {Paris}}}]{GFKP-5pts:2014mn}
\bibinfo{author}{\bibfnamefont{E.}~\bibnamefont{{Grohs}}},
  \bibinfo{author}{\bibfnamefont{G.~M.} \bibnamefont{{Fuller}}},
  \bibinfo{author}{\bibfnamefont{C.~T.} \bibnamefont{{Kishimoto}}},
  \bibnamefont{and} \bibinfo{author}{\bibfnamefont{M.~W.}
  \bibnamefont{{Paris}}}, \bibinfo{journal}{\jcap}
  \textbf{\bibinfo{volume}{5}}, \bibinfo{pages}{17} (\bibinfo{year}{2015}),
  \eprint{1502.02718}.

\bibitem[{\citenamefont{{Fuller} and {Kishimoto}}(2009)}]{2009PhRvL.102t1303F}
\bibinfo{author}{\bibfnamefont{G.~M.} \bibnamefont{{Fuller}}} \bibnamefont{and}
  \bibinfo{author}{\bibfnamefont{C.~T.} \bibnamefont{{Kishimoto}}},
  \bibinfo{journal}{Phys.~Rev.~Lett.} \textbf{\bibinfo{volume}{102}},
  \bibinfo{eid}{201303} (\bibinfo{year}{2009}), \eprint{0811.4370}.

\bibitem[{\citenamefont{{Dolgov} et~al.}(1990)\citenamefont{{Dolgov}, {Sazhin},
  and {Zeldovich}}}]{1990bmc..book.....D}
\bibinfo{author}{\bibfnamefont{A.~D.} \bibnamefont{{Dolgov}}},
  \bibinfo{author}{\bibfnamefont{M.~V.} \bibnamefont{{Sazhin}}},
  \bibnamefont{and} \bibinfo{author}{\bibfnamefont{Y.~B.}
  \bibnamefont{{Zeldovich}}}, \emph{\bibinfo{title}{{Basics of modern
  cosmology}}} (\bibinfo{year}{1990}).

\bibitem[{\citenamefont{{Izotov} and {Thuan}}(2010)}]{Izotov:2010ns}
\bibinfo{author}{\bibfnamefont{Y.~I.} \bibnamefont{{Izotov}}} \bibnamefont{and}
  \bibinfo{author}{\bibfnamefont{T.~X.} \bibnamefont{{Thuan}}},
  \bibinfo{journal}{\apjl} \textbf{\bibinfo{volume}{710}}, \bibinfo{pages}{L67}
  (\bibinfo{year}{2010}).

\bibitem[{\citenamefont{{Aver} et~al.}(2013)\citenamefont{{Aver}, {Olive},
  {Porter}, and {Skillman}}}]{2013JCAP...11..017A}
\bibinfo{author}{\bibfnamefont{E.}~\bibnamefont{{Aver}}},
  \bibinfo{author}{\bibfnamefont{K.~A.} \bibnamefont{{Olive}}},
  \bibinfo{author}{\bibfnamefont{R.~L.} \bibnamefont{{Porter}}},
  \bibnamefont{and} \bibinfo{author}{\bibfnamefont{E.~D.}
  \bibnamefont{{Skillman}}}, \bibinfo{journal}{\jcap}
  \textbf{\bibinfo{volume}{11}}, \bibinfo{eid}{017} (\bibinfo{year}{2013}),
  \eprint{1309.0047}.

\bibitem[{\citenamefont{{Dolgov} et~al.}(2002)}]{2002NuPhB.632..363D}
\bibinfo{author}{\bibfnamefont{A.~D.} \bibnamefont{{Dolgov}}}
  \bibnamefont{et~al.}, \bibinfo{journal}{Nucl. Phys. B}
  \textbf{\bibinfo{volume}{632}}, \bibinfo{pages}{363} (\bibinfo{year}{2002}),
  \eprint{hep-ph/0201287}.

\bibitem[{\citenamefont{Abazajian et~al.}(2015)}]{Abazajian201566}
\bibinfo{author}{\bibfnamefont{K.}~\bibnamefont{Abazajian}}
  \bibnamefont{et~al.}, \bibinfo{journal}{Astroparticle Physics}
  \textbf{\bibinfo{volume}{63}}, \bibinfo{pages}{66 } (\bibinfo{year}{2015}),
  ISSN \bibinfo{issn}{0927-6505}.

\bibitem[{\citenamefont{{The Polarbear Collaboration: P.~A.~R.~Ade}
  et~al.}(2014)}]{2014ApJ...794..171T}
\bibinfo{author}{\bibnamefont{{The Polarbear Collaboration: P.~A.~R.~Ade}}}
  \bibnamefont{et~al.}, \bibinfo{journal}{\apj} \textbf{\bibinfo{volume}{794}},
  \bibinfo{eid}{171} (\bibinfo{year}{2014}), \eprint{1403.2369}.

\bibitem[{\citenamefont{{Carlstrom} and {Spt
  Collaboration}}(2003)}]{SPT:2003IA}
\bibinfo{author}{\bibfnamefont{J.~E.} \bibnamefont{{Carlstrom}}}
  \bibnamefont{and} \bibinfo{author}{\bibnamefont{{Spt Collaboration}}},
  \bibinfo{journal}{IAU Special Session} \textbf{\bibinfo{volume}{2}},
  \bibinfo{pages}{34} (\bibinfo{year}{2003}).

\bibitem[{\citenamefont{{Abazajian} et~al.}(2005)\citenamefont{{Abazajian},
  {Bell}, {Fuller}, and {Wong}}}]{2005PhRvD..72f3004A}
\bibinfo{author}{\bibfnamefont{K.}~\bibnamefont{{Abazajian}}},
  \bibinfo{author}{\bibfnamefont{N.~F.} \bibnamefont{{Bell}}},
  \bibinfo{author}{\bibfnamefont{G.~M.} \bibnamefont{{Fuller}}},
  \bibnamefont{and} \bibinfo{author}{\bibfnamefont{Y.~Y.~Y.}
  \bibnamefont{{Wong}}}, \bibinfo{journal}{\prd} \textbf{\bibinfo{volume}{72}},
  \bibinfo{eid}{063004} (\bibinfo{year}{2005}), \eprint{astro-ph/0410175}.

\bibitem[{\citenamefont{{Kishimoto} et~al.}(2006)\citenamefont{{Kishimoto},
  {Fuller}, and {Smith}}}]{2006PhRvL..97n1301K}
\bibinfo{author}{\bibfnamefont{C.~T.} \bibnamefont{{Kishimoto}}},
  \bibinfo{author}{\bibfnamefont{G.~M.} \bibnamefont{{Fuller}}},
  \bibnamefont{and} \bibinfo{author}{\bibfnamefont{C.~J.}
  \bibnamefont{{Smith}}}, \bibinfo{journal}{Phys.~Rev.~Lett.}
  \textbf{\bibinfo{volume}{97}}, \bibinfo{eid}{141301} (\bibinfo{year}{2006}),
  \eprint{astro-ph/0607403}.

\bibitem[{\citenamefont{{Hamann} et~al.}(2010)\citenamefont{{Hamann},
  {Hannestad}, {Raffelt}, {Tamborra}, and {Wong}}}]{2010PhRvL.105r1301H}
\bibinfo{author}{\bibfnamefont{J.}~\bibnamefont{{Hamann}}},
  \bibinfo{author}{\bibfnamefont{S.}~\bibnamefont{{Hannestad}}},
  \bibinfo{author}{\bibfnamefont{G.~G.} \bibnamefont{{Raffelt}}},
  \bibinfo{author}{\bibfnamefont{I.}~\bibnamefont{{Tamborra}}},
  \bibnamefont{and} \bibinfo{author}{\bibfnamefont{Y.~Y.~Y.}
  \bibnamefont{{Wong}}}, \bibinfo{journal}{Phys.~Rev.~Lett.}
  \textbf{\bibinfo{volume}{105}}, \bibinfo{eid}{181301} (\bibinfo{year}{2010}),
  \eprint{1006.5276}.

\bibitem[{\citenamefont{{Hamann} et~al.}(2011)\citenamefont{{Hamann},
  {Hannestad}, {Raffelt}, and {Wong}}}]{2011JCAP...09..034H}
\bibinfo{author}{\bibfnamefont{J.}~\bibnamefont{{Hamann}}},
  \bibinfo{author}{\bibfnamefont{S.}~\bibnamefont{{Hannestad}}},
  \bibinfo{author}{\bibfnamefont{G.~G.} \bibnamefont{{Raffelt}}},
  \bibnamefont{and} \bibinfo{author}{\bibfnamefont{Y.~Y.~Y.}
  \bibnamefont{{Wong}}}, \bibinfo{journal}{\jcap} \textbf{\bibinfo{volume}{9}},
  \bibinfo{eid}{034} (\bibinfo{year}{2011}), \eprint{1108.4136}.

\bibitem[{\citenamefont{{Dvorkin} et~al.}(2014)\citenamefont{{Dvorkin},
  {Wyman}, {Rudd}, and {Hu}}}]{2014PhRvD..90h3503D}
\bibinfo{author}{\bibfnamefont{C.}~\bibnamefont{{Dvorkin}}},
  \bibinfo{author}{\bibfnamefont{M.}~\bibnamefont{{Wyman}}},
  \bibinfo{author}{\bibfnamefont{D.~H.} \bibnamefont{{Rudd}}},
  \bibnamefont{and} \bibinfo{author}{\bibfnamefont{W.}~\bibnamefont{{Hu}}},
  \bibinfo{journal}{\prd} \textbf{\bibinfo{volume}{90}}, \bibinfo{eid}{083503}
  (\bibinfo{year}{2014}), \eprint{1403.8049}.

\bibitem[{\citenamefont{{Cirelli} et~al.}(2005)\citenamefont{{Cirelli},
  {Marandella}, {Strumia}, and {Vissani}}}]{2005NuPhB.708..215C}
\bibinfo{author}{\bibfnamefont{M.}~\bibnamefont{{Cirelli}}},
  \bibinfo{author}{\bibfnamefont{G.}~\bibnamefont{{Marandella}}},
  \bibinfo{author}{\bibfnamefont{A.}~\bibnamefont{{Strumia}}},
  \bibnamefont{and}
  \bibinfo{author}{\bibfnamefont{F.}~\bibnamefont{{Vissani}}},
  \bibinfo{journal}{Nuclear Physics B} \textbf{\bibinfo{volume}{708}},
  \bibinfo{pages}{215} (\bibinfo{year}{2005}), \eprint{hep-ph/0403158}.

\end{thebibliography}

\end{document}